\documentstyle[prl,aps,preprint]{revtex}
\begin{document}
\draft
%
\title{A Semiclassical Calculation of Scars for a Smooth Potential}
\author{Daniel Provost\cite{byline} and Michel Baranger}
\address{
Center for Theoretical Physics,
Laboratory for Nuclear Science,
and Department of Physics,
Massachusetts Institute of Technology,
Cambridge, Massachusetts\ \ 02139\ \ \ U.S.A.
}
%
\date{January 13, 1993}
\maketitle
\begin{abstract}
Bogomolny's formula for energy--smoothed
scars is applied for the first
time to a non--specific, non--scalable Hamiltonian,
a two--dimensional anharmonic oscillator.
The semiclassical theory reproduces well
the exact quantal results
over a large spatial and energy range.
\end{abstract}
\pacs{PACS numbers: $03.65.$Sq, $05.45.+$b }

\narrowtext
The scars left on stationary state
wavefunctions by unstable classical
periodic orbits were discovered by
McDonald and Kaufman\cite{MK78}.
They were named and further studied by Heller\cite{He84}.
The paradox of the scars is that they occur
in an energy region where
a generic classical trajectory covers the energy shell uniformly;
in spite of this, many quantal wavefunctions like to concentrate in the
vicinity of one or several periodic orbits.
Obviously, an understanding of this paradox requires
a semi-classical theory.

Such a semi-classical theory of scars in the usual, coordinate space
wave functions was given by Bogomolny\cite{Bo88}.
There are also semi-classical
theories of scars in the Wigner distribution\cite{Be89} and in the
Husimi distribution\cite{ABpre}.
We present here some results from an extensive
comparison of the exact quantal scars with
the Bogomolny scars\cite{Bo88} for an ordinary smooth potential.
Apart from an interesting qualitative discussion\cite{WH89},
we do not know of any other detailed comparison for a sufficiently
``generic'' Hamiltonian.

The famous Gutzwiller trace formula\cite{Gu71}
can be obtained by integrating the
Bogomolny formula over the space coordinates.  Therefore our work leads
also to a check of the Gutzwiller trace formula for our Hamiltonian.
There have already been a few checks of similar quality for the
Gutzwiller trace formula with general enough Hamiltonians,
for instance Ref.~\cite{Wi88}.

Our Hamiltonian is
\begin{equation}
H(p_x,p_y,x,y)={1\over 2}(p_x^2+p_y^2)+0.05x^2+ (y-x^2/2)^2.\label{eham}
\end{equation}
Its classical dynamics and periodic orbits have been studied in
detail\cite{BD87}.
We had to continue the study of periodic orbits towards higher
energy, using a totally new method for which there is no space here.
Suffice it to say  that this method ensures that no orbit below
period $20$   is missed.
Fig.~\ref{fpot}
shows a contour plot of the potential $V(x,y)=H-{1\over 2}p^2$
for energies of interest.
The transition from mostly regular to mostly
chaotic motion happens for energies of order 0.1.
Islands of regularity remain, however, no matter how high in energy
one goes.
But at the energies of our numerical comparisons,
they are very tiny.

We took $\hbar=0.05$.
The calculation of the exact quantal wave functions
and energies was done with the basis
$\phi_m^{(x)}(x)\phi_n^{(y)}(y-x^2/2)$,
where $\phi_m^{(x)}$    and $\phi_n^{(y)}$
are harmonic oscillator wave functions appropriate for
the bottom of the well.
We used 240 oscillator states for the x direction
and 26 for y.
A sensitive test of the sufficiency of this basis comes from a
comparison with the smooth Thomas--Fermi density of states,
including the corrections\cite{BTUpre} of order $\hbar^2$.
The test shows that our basis
begins to fail for an energy somewhere between
$0.80$ and $0.85$,
and consequently we stopped our comparisons there.

Bogomolny's formula is\cite{Bo88}
\begin{eqnarray}
\Delta({\bf q},E;\epsilon)
&&= {2\over(2\pi\hbar)^{3/2}}\sum_{\rm periodic \atop orbits}
  \tilde f_{\epsilon}(\tau){1\over \vert\dot q_1\vert}
  {1\over\sqrt{\vert m_{qp}\vert}}\nonumber\\
&&\times \cos\Bigl[{1\over\hbar}\Bigl(\bar S +{1\over 2}
   {tr{\cal M}-2\over m_{qp}} q_2^2\Bigr)
   - {\pi\over 2}\mu-{\pi\over 4}\Bigr].\label{ebogo}
\end{eqnarray}
$\Delta$ is the oscillating part
(as explained in the next sentence) of the
energy--smoothed coordinate space probability density
\begin{equation}
\sum_n f_\epsilon(E-E_n)\,\,
\vert\!\!<{\bf q}\vert\psi_n>\!\!\vert^2 ,\label{eexact}
\end{equation}
where $f_{\epsilon}$ is the smoothing function,
which we take to be the
normalized gaussian
\begin{equation}
f_\epsilon(E)=(2\pi)^{-1/2} \epsilon^{-1} e^{-E^2/2\epsilon^2}.
\label{egaus}
\end{equation}
$\Delta$ is obtained by subtracting from Eq.~(\ref{eexact})
the energy--smoothed Thomas--Fermi
density.
On the right hand side of Bogomolny's formula,
$\tilde f_\epsilon(\tau)= e^{-\epsilon^2\tau^2/2\hbar^2}$ is the
Fourier transform of $f_\epsilon(E)$, $\tau$ is the period
of the  periodic orbit,
and $\bar S$ is its action $\int {\bf p}\cdot d{\bf q}$.
The coordinates $q_1$ and $q_2$
are chosen especially for each periodic orbit,
$q_1$ being the distance along the orbit and $q_2$ being the
perpendicular coordinate.
$\cal M$ is the 2\,x\,2 submatrix of the monodromy
matrix involving  coordinates $q_2$ and $p_2$,
and $m_{qp}$ is one of
its off-diagonal elements.
Finally $\mu$ is equal to $\mu_m$, the Maslov index of the
orbit\cite{CRG90},
when $m_{qp}$ and ${\rm Tr} {\cal M}-2$ have the same sign;
$\mu$ is equal to $\mu_m-1 $ when the signs are opposite.

In order that Eq.~(\ref{ebogo})
be valid, it is also necessary to perform
some smoothing over coordinate space on both sides of the equation.
There are two reasons for this: (a) it is essential for reducing
the semiclassical contribution to a sum over periodic
orbits\cite{Bo88};
(b) the semiclassical theory is
not valid near the points where $m_{qp}=0$,
which are the self-conjugate points,
and the spatial smoothing minimizes this discrepancy.
We smoothed with a gaussian proportional to
$e^{-(q_x^2+q_y^2)/b^2}$.
This is equivalent to calculating the coordinate space projection
of the Husimi distribution.
We chose $b=0.2$; more about this later.

Gutzwiller's trace formula\cite{Gu71},
obtained by integrating Eq.~(\ref{ebogo}) over space, is
\begin{equation}
d(E;\epsilon)={1\over\pi\hbar}
\!\! \sum_{\rm periodic \atop orbits}
\!\! \tilde f_\epsilon(\tau)
{\tau_o\over\sqrt{\vert tr{\cal M}-2\vert}}
\cos( {\bar S\over \hbar} -{\pi\over 2}\mu_m ).\label{egutz}
\end{equation}
$d$ is the oscillating part of the energy--smoothed density of states,
calculated by subtracting the Thomas--Fermi density of states and its
corrections\cite{BTUpre}
of order $\hbar^2$. For orbits consisting of repeated traversals
of a primitive periodic orbit, $\tau_o$ is the period of the latter.

It has long been everybody's dream to get rid of the energy smoothing;
to let $\epsilon \rightarrow 0$ and to use
Eqs.~(\ref{ebogo}) and~(\ref{egutz})
to predict individual stationary states.
Like everyone else we avoided this limit, as the
process seems to diverge.
Instead, we chose our $\epsilon$ so that
relatively few periodic orbits would contribute.

Fig.~\ref{fper}
shows pictures of the first 12 orbits in order of increasing period,
again at $E=0.8$.
We have worked in the range
$0.5\leq E\leq 0.85$ and , for $E\neq 0.8$,
the order might be slightly different from that in Fig.~\ref{fper}.
In the distribution of periods, the three lowest
$(4.44,6.44,7.14)$
are clearly separated from those above
$(10.51,11.57,11.60, \ldots etc.$).
Hence we expect that, by choosing $\epsilon$ large enough,
many features can
be described in terms of 3 periodic orbits only.
This turns out to be true indeed.
Fig.~\ref{fdens}
shows the oscillations
in the smoothed energy level density, as given by the Gutzwiller
formula, Eq.~(\ref{egutz}),
calculated with $\epsilon=0.01$ and 5 periodic orbits,
but it is only very slightly better
than that calculated with 3 orbits.
On the other hand it is radically different
from that calculated with only
1 or 2 orbits.
In the case of 3 or more orbits, the agreement with the
exact quantal calculation can be termed
very good for this value of $\epsilon$.

Fig.~\ref{fstate}
shows the most striking scar we have found: odd (in $x$) state
no. 145, with $E=0.814$,
scarred by the third periodic orbit, a simple asymmetric libration.
Our calculation of energy--smoothed scars according to the Bogomolny
formula, Eq.~(\ref{ebogo}),
is illustrated in
Figs.~\ref{fcompa},~\ref{fcompb}, and~\ref{fcompc}
and compared with the exact result.
For lack of space, we concentrate here on the energy range
$0.726\leq E\leq 0.814$, which is typical of other energies.
Fig.~\ref{fcompa} is for the same energy as Fig.~\ref{fstate},
and the scar is still prominent.
Figs.~\ref{fcompb} and~\ref{fcompc} are two other examples.
They all show some combination of scarring by the first three orbits.
The agreement between the exact quantal and
the semi--classical pictures
can be described as following the main trends
very well, but quantitatively inaccurate.
By ``main trends'', we mean in particular the way that the density
fluctuation in the vicinity
of each orbit (the ``scar'' of that orbit)
oscillates and changes sign as a function of energy.
This is better seen in Fig.~\ref{fpoint}, which shows the scar
strength (both exact and semiclassical) as a function of energy for
three points in the $xy$ plane.

The agreement between Eq.~(\ref{ebogo}) and the exact quantal density
is limited to the central part of our potential, exclusive of the
two ``arms'' (see Fig.~\ref{fpot}).
This is because, in the arms, the motion is approximately integrable,
consisting of fast transverse oscillations whose action is an
adiabatic invariant for the slower longitudinal motion\cite{EHP89}.
As a result,
the exact density in the arms exhibits a very simple pattern,
visible in the upper corners of
Figs.~\ref{fcompa},~\ref{fcompb}, and~\ref{fcompc},
and consisting mostly of wide
longitudinal oscillations whose wavelength
can be calculated very simply with the above adiabatic picture.
The short--period classical orbits, on the other hand, are found only
in the central, chaotic region.
It is possible to increase the value of $b$,
the spatial smoothing width,
so as to wash out the oscillations in the arms,
but then this would also
wash out many interesting features in the center.
The value chosen, $b=0.2$, is a reasonable compromise.

\acknowledgments

This work was supported in part by funds
provided by the U. S. Department of Energy (D.O.E.) under contract
\#DE-AC02-76ER03069.
\begin{figure}
\caption{Three contours of the potential $V(x,y)$: $E=0.1$ (dots),
$E=0.4$ (dashes), $E=0.8$ (full).}
\label{fpot}
\end{figure}
\begin{figure}
\caption{The 12 orbits with lowest periods for $E=0.8$.}
\label{fper}
\end{figure}
\begin{figure}
\caption{The smoothed energy level density,
with $\epsilon=0.01$, minus the
Thomas--Fermi density.
The full line is the exact quantal
calculation, the dashed line is
Eq.~(5) with 5 orbits.
The Thomas--Fermi density at $E=0.65$ is $\simeq 290$.
Hence the average number of stationary states inside
one $\epsilon$ is $2.9$.}
\label{fdens}
\end{figure}
\begin{figure}
\caption{Contours of the exact
coordinate space probability density for
odd state no. 145,
showing striking scarring by the no. 3 periodic orbit
and its mirror reflection $x \rightarrow -x$.}
\label{fstate}
\end{figure}
\begin{figure}
\caption{The oscillating part , $\Delta$, of the
en\-er\-gy--\-and\---\-spa\-ce\---\-smoo\-thed
coordinate space density.
Energy smoothing parameter $\epsilon=0.01$.
Top: exact quantal.
Bottom: Eq.~(2) with 5 orbits.
Solid contours are positive, dashed are negative. The contour
spacing is 0.5 in all
Figs.~5,~6,
and~7.
Contour 0 is not drawn.
The Thomas--Fermi density is $1/2\pi\hbar^2=63.7$. The energy is
$E=0.814$.}
\label{fcompa}
\end{figure}
\begin{figure}
\caption{Same as
Fig.~5
for $E=0.794$.}
\label{fcompb}
\end{figure}
\begin{figure}
\caption{Same as
Fig.~5
for $E=0.746$.}
\label{fcompc}
\end{figure}
\begin{figure}
\caption{The exact scar strength $\Delta$ (full line)
and the semi--classical one calculated with 5 orbits (dashed),
as a function of energy, at three points of coordinate space.
Bottom: $x=0$, $y=-0.65$, emphasizing orbit no.\ 1.
Middle: $x=0.8$, $y=0$, emphasizing orbit no.\ 2.
Top: $x=0.75$, $y=0.75$, emphasizing orbit no.\ 3.}
\label{fpoint}
\end{figure}

\begin{references}
%
\bibitem[*]{byline}Present Address:
Chemical Physics Theory Group,
Department of Chemistry,
University of Toronto,
Toronto, Canada~~M5S~1A1
%
\bibitem{MK78}S. W. McDonald and A. N. Kaufman,
Phys.\ Rev.\ Lett.\ {\bf 42}, 1189 (1979).
%
\bibitem{He84}E. J. Heller,
Phys.\ Rev.\ Lett.\ {\bf 53}, 1515 (1984).
%
\bibitem{Bo88}E. B. Bogomolny,
Physica D {\bf 31}, 169 (1988).
%
\bibitem{Be89}M. V. Berry,
Proc.\ R. Soc.\ London {\bf A423 }, 219 (1989).
%
\bibitem{ABpre}M. A. M. deAguiar and M. Baranger,
(to be published).
%
\bibitem{WH89}D. Wintgen and A. H\"onig,
Phys.\ Rev.\ Lett. {\bf 63}, 1467 (1989).
%
\bibitem{Gu71}M. C. Gutzwiller,
J.\ Math.\ Phys.\ {\bf 12}, 343 (1971).
%
\bibitem{Wi88}D. Wintgen,
Phys.\ Rev.\ Lett. {\bf 61}, 1803 (1988).
%
\bibitem{BD87}M.\ Baranger and K.\ T.\ R.\ Davies,
Ann.\ Phys.\ (NY) {\bf 177}, 330 (1987).
%
\bibitem{BTUpre}A. Bohigas, S. Tomsovic, and D. Ullmo,
Phys.\ Rep.\ (in press).
%
\bibitem{CRG90}
S. C. Creagh, J. M. Robbins, and R. G. Littlejohn,
Phys.\ Rev.\ A {\bf 42}, 1907 (1990);
J. M. Robbins,
Nonlinearity {\bf 4}, 343 (1991).
%
%
\bibitem{EHP89}B. Eckhardt, G. Hose, and E. Pollak,
Phys.\ Rev.\ A {\bf 39}, 3776 (1989).
%
\end{references}
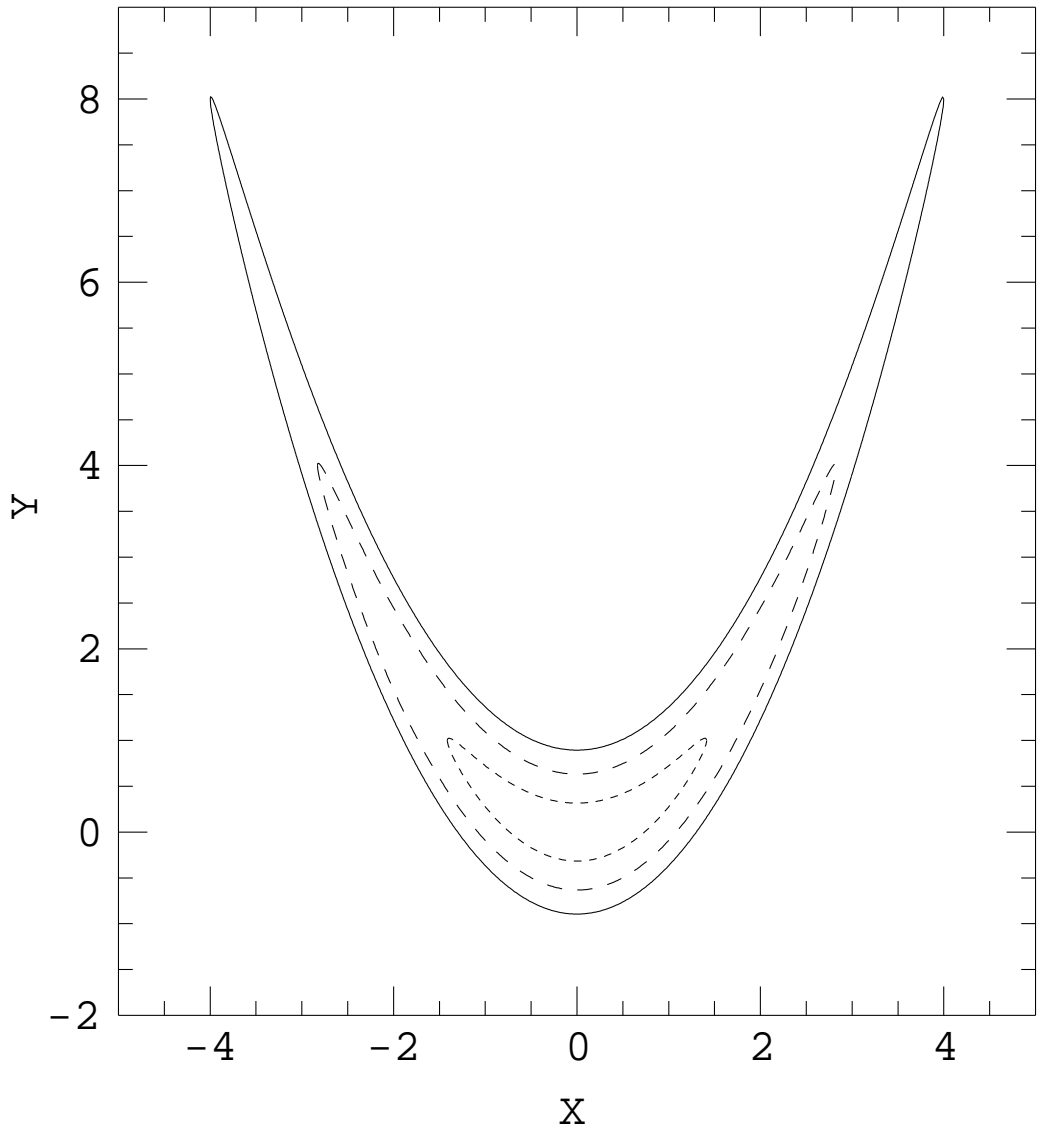
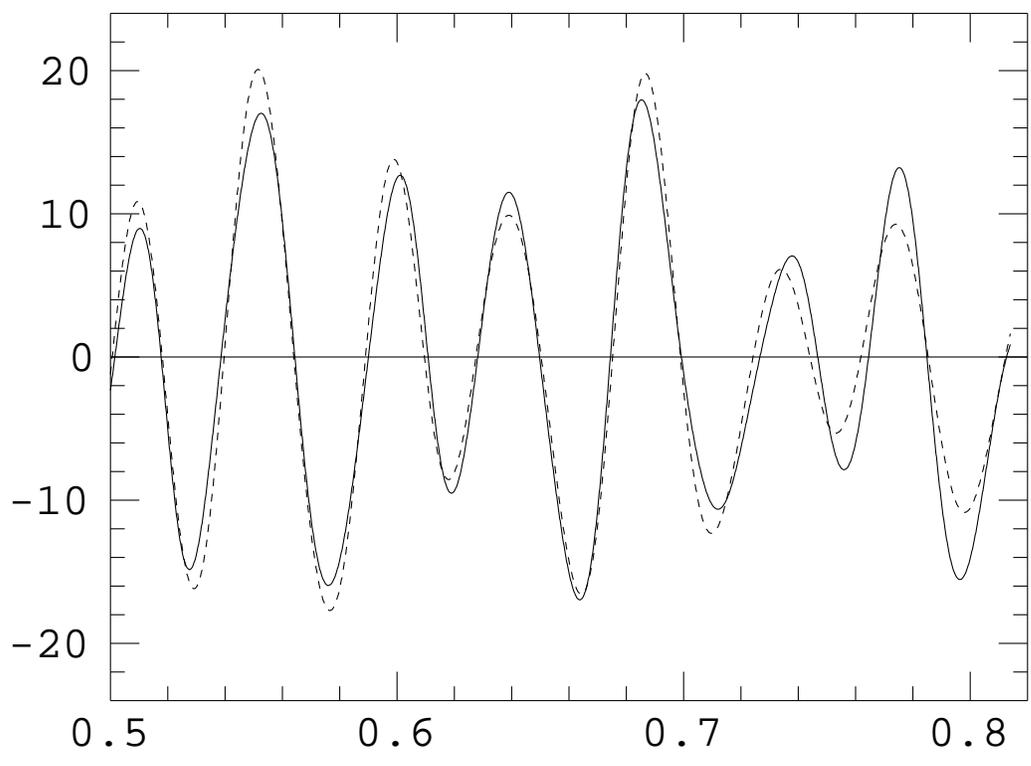
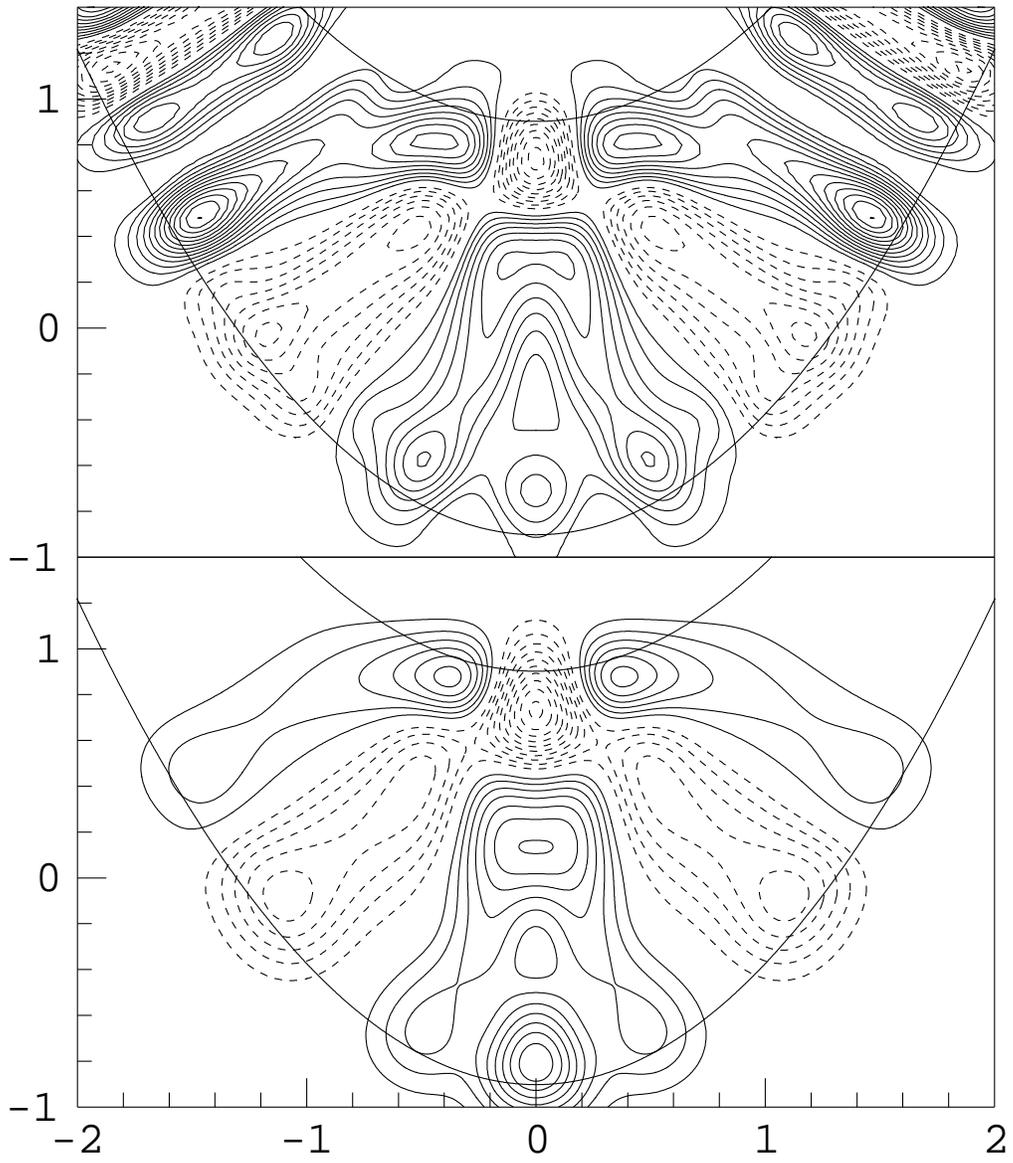
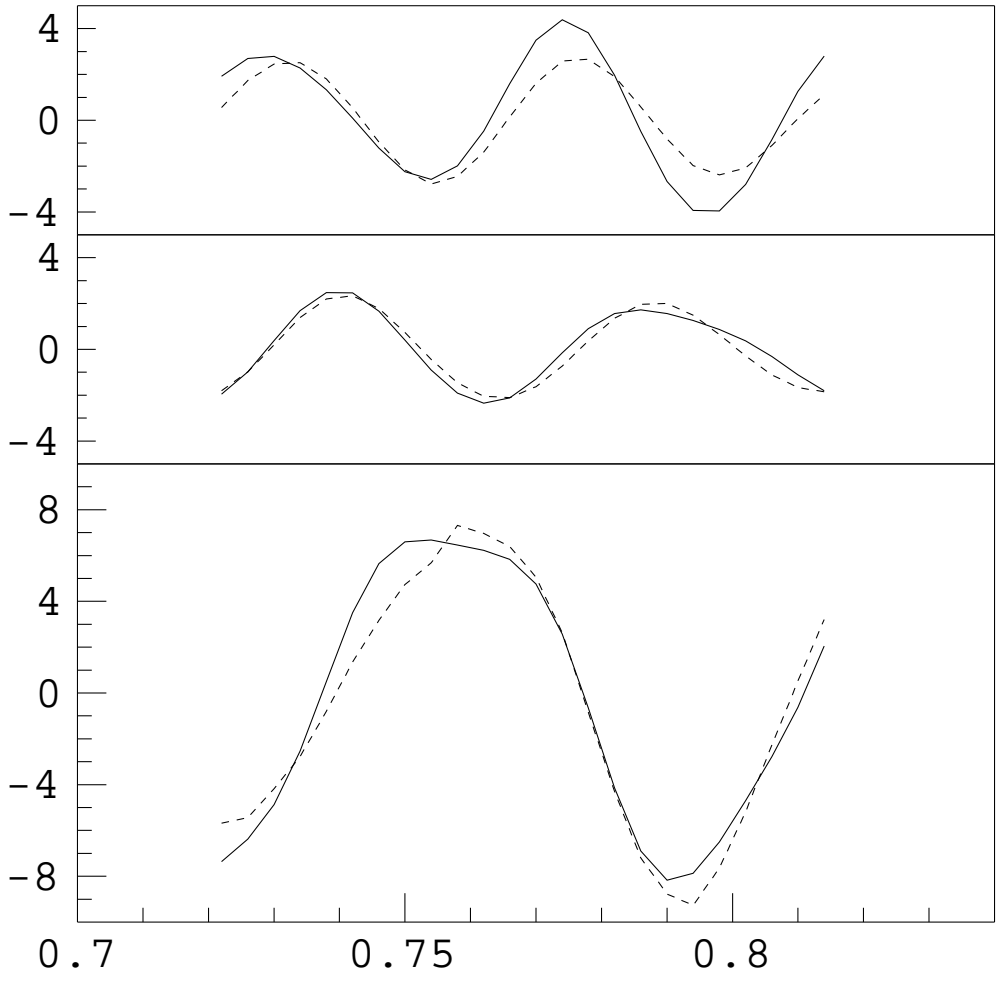
\end{document}